# DISTRIBUTED POWER SUPPLY CONTROL AND MONITORING SYSTEM

J. M. Nogiec, E. Desavouret, FNAL*, Batavia, IL60510, USA

## Abstract

An extensible power supply control and monitoring system has been developed at Fermilab's Technical Division to control and monitor power supplies of various types from within many different applications. This system, deployed as a network service, provides uniform programming and user interfaces for various types of power supplies, ranging from 20A to 30kA.

## 1 SYSTEM ORGANIZATION

The Magnet Test Facility at Fermilab uses many different power supplies in accomplishing its mission. This variety requires the availability of transparent access to these power supplies for applications and users alike. To satisfy this requirement, a power supply service has been developed that allows for concurrent access to a single power supply by several client programs. Only one program is granted control capabilities at any given time (other clients may run in the monitoring mode). For example, several users can monitor the current changes requested by an automated test program. Additionally, the distributed architecture of the system allows users and system operators to be geographically dispersed (see Fig. 1).

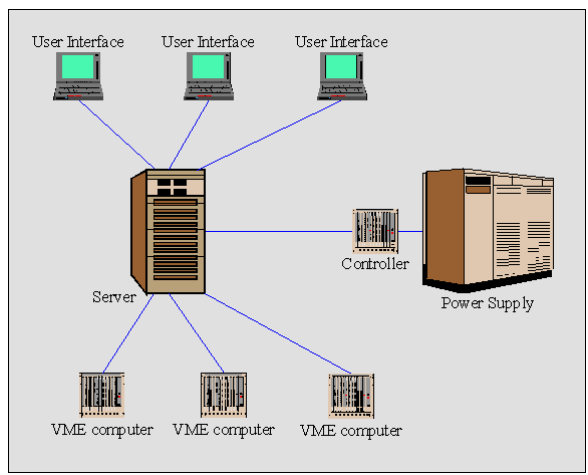

Figure 1: Architecture of the power control system.

The system is based on the server-client model. Each power supply has a central server that is connected via a dedicated fiber optic link to an embedded power supply controller [1]. Clients execute power control access operations that are converted into request messages and sent to the power supply server. The server receives the request, unpacks it, and sends a message to the specific power supply controller that performs the actual operation, then waits for the reply from the controller and sends the response back to the client. Before a client can control the power supply it must allocate and lock the control of this device.

Supported server platforms include Solaris and VxWorks, and client platforms include VxWorks, Solaris, and Java.

## 2 RAMP PROFILES

Accelerator physics applications as well as the inductive nature of the load dictate complex waveforms of the magnet current. To cope with this requirement the system is equipped with a proprietary ramp definition language that can be used to define sophisticated ramp profiles.

The language consists of the following six statements:

- *ramp*: ramp the current until the destination current is reached with a given initial ramping speed and acceleration.
- *smooth ramp*: execute a parabolic ramp between the starting and destination currents with a given maximum ramping speed and acceleration distance.
- *delay*: pause for a requested time.
- *trigger*: issue an external trigger signal,
- *repeat*: execute a set of statements a given number of times.
- *loop*: repeat a set of statements until some condition is satisfied.

A ramp profile program consists of a sequence of these statements and C-style comments. A simple profile to ramp current up and down using a standard parabolic profile is shown in Figure 2.

```
/* Parabolic ramp profile */
ramp 0, -20, 0
repeat 2
   smooth_ramp 0, 1000, 200, 20
   trigger
   smooth_ramp 1000, 0, 200, 20
   delay 5000
end
```

Figure 2: Parabolic ramp profile.

A resulting current trace is shown in Figure 3.

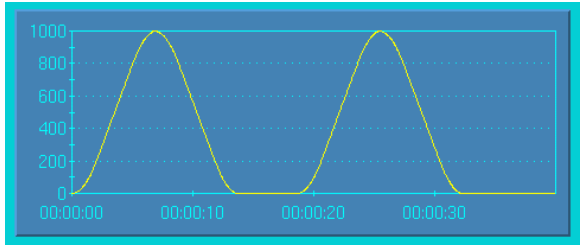

Figure 3: Current trace.

A more complicated profile to degauss magnets is shown in Figure 4.

```
/* Degaussing ramp profile */
ramp 0, -100, 0
delay 20
ramp 1.356, 2.71248, 0.0
ramp 2.516, 2.31958, 0.0
ramp 3.378, 1.72298, 0.0
ramp 3.872, 0.98992, 0.0
ramp 3.971, 0.19752, 0.0
ramp 3.684, -0.57524, 0.0
ramp 3.056, -1.25502, 0.0
....
ramp -0.004,  0.00672, 0.0
ramp -0.000,  0.00706, 0.0
```

Figure 4: Excerpts from a degaussing ramp profile.

Ramp profiles can be previewed before execution with help of a graphical simulator (see Fig. 5).

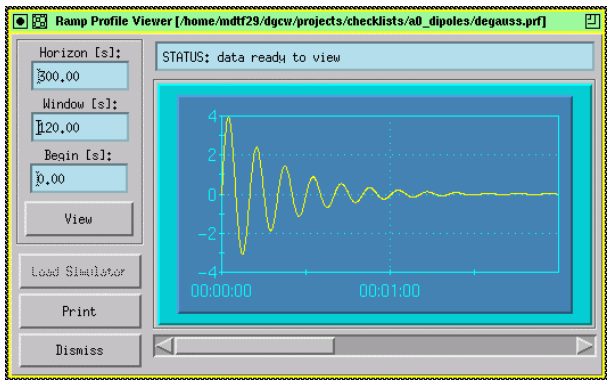

Figure 5: The simulator showing a degaussing profile.

Successfully tested profiles can be loaded and executed on the power supply controller using either the graphical user interface or a dedicated profile loading program. Current traces from the execution of ramp profiles can be captured with help of a dedicated monitoring program.

## 3 API

A C language API that has been developed for the VxWorks and Unix platforms allows user programs to control power supplies. A functionally equivalent Java API has also been created. Each of these APIs provides the user with over 30 methods that allow applications to specify protocol parameters, load various ramp segments into the memory of the power supply controller, control the execution of ramp profiles, control the current trace capture mechanism, examine uploaded profiles, read input and output voltages proportional to the drive signal and output current, establish operation mode, and directly control the power supply state.

## 4 SCRIPTING

Many specialized client programs have been developed, including GUI, test systems, monitoring applications, current tracker application, R/T monitoring and control systems [3], and measurement systems [2][4]. Additionally, a set of programs specifically designed for use in scripts has been provided for the user. These programs allow for ramping current, reading current, waiting for the completion of the ramp, loading and executing ramp profiles, examining the status of the power supply and controller, waiting on a specific status of the controller or power supply, and identifying the power supply.

## 5 INTERACTIVE CONTROL

The system provides users with graphical user interfaces to visualize current history (see Fig. 6), power supply status information (see Fig. 7), and to control power supplies from remote locations. These interfaces allow for monitoring of other power control programs, charting of current trends, switching and resetting power supplies,  and  taking control from run-away applications. In addition, ramp operations that are in progress can be stopped or suspended. Currently, the system provides graphical operator consoles for the X-Windows and Java platforms.

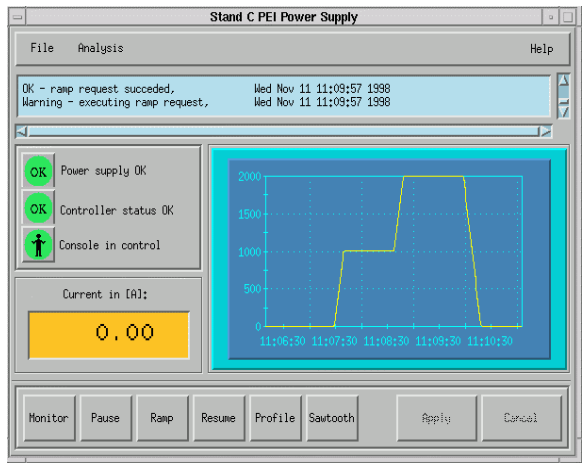

Figure 6: X-Windows power supply control interface.

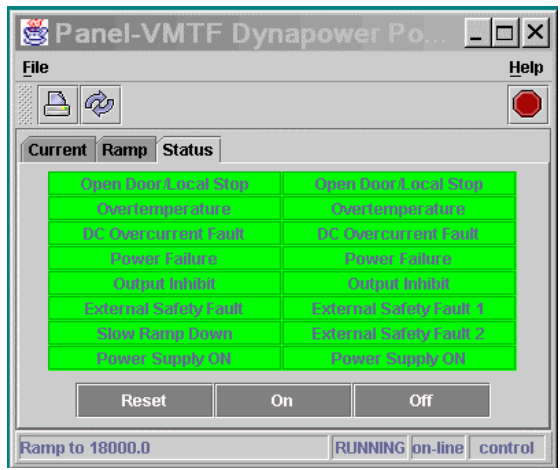

Figure 7: Power supply status in a Java application.

## 6 SUMMARY

The presented power supply control system allows client programs to control and monitor power supplies over the network. The system allows for concurrent multi-user access, with at most one application at a time granted controlling capabilities. Users can control and monitor the state of a power supply as well as monitor generated current by:

- Sending separate requests to the service via provided C/C++ and Java APIs.
- Uploading current control programs written in a specialized ramp profile definition language to be interpreted by the system.
- Scripting using a provided set of programs.
- Using supplied GUIs.

In addition, special programs are available to monitor and log all control requests being sent to the service and to provide traces of current. Also, a graphical simulator is provided to facilitate testing of ramp profiles developed by the user.

The power control and monitoring system has been deployed at Fermilab's Magnet Test Facility to provide access to power supplies of eight different types. This system can be easily extended by adding new power supply types and new client programs.

## REFERENCES

[1] S. Sharonov, J.M. Nogiec, "An Embedded Power Supply Controller", PAC'97, Vancouver, 1997.
[2] J.M. Nogiec et al, "A Flexible and Configurable System to Test Accelerator Magnets", PAC'01, Chicago, 2001.
[3] J.M. Nogiec et al, "A Distributed Monitoring and Control System", PAC'97, Vancouver, 1997.
[4] J.W. Sim et al, "A Software System for Measurement of Accelerator Magnets Using a Relational Database", MT-16, Tallahassee, Florida, 1999.